\documentclass[aps,pra,twocolumn,letterpaper,superscriptaddress]{revtex4-1}
\usepackage[utf8x]{inputenc}
\usepackage[T1]{fontenc}
\usepackage{mathptmx}

\usepackage{mathtools}
\usepackage[binary-units]{siunitx}
\usepackage{hyperref}
\usepackage[capitalise]{cleveref}
\usepackage{xfrac}

\usepackage{graphicx}
\usepackage{multirow}
\usepackage{longtable}
\usepackage{booktabs}
\usepackage{color}

\begin{document}

\title{Experimental quantum key distribution with simulated ground-to-satellite photon losses and processing limitations}
\author{Jean-Philippe Bourgoin}
\email{jbourgoin@uwaterloo.ca}
\affiliation{Institute for Quantum Computing, University of Waterloo, Waterloo, ON N2L 3G1, Canada}
\affiliation{Department of Physics and Astronomy, University of Waterloo, Waterloo, ON N2L 3G1, Canada}
\author{Nikolay Gigov}
\affiliation{Institute for Quantum Computing, University of Waterloo, Waterloo, ON N2L 3G1, Canada}
\affiliation{Department of Physics and Astronomy, University of Waterloo, Waterloo, ON N2L 3G1, Canada}
\author{Brendon~L. Higgins}
\affiliation{Institute for Quantum Computing, University of Waterloo, Waterloo, ON N2L 3G1, Canada}
\affiliation{Department of Physics and Astronomy, University of Waterloo, Waterloo, ON N2L 3G1, Canada}
\author{Zhizhong Yan}
\affiliation{Institute for Quantum Computing, University of Waterloo, Waterloo, ON N2L 3G1, Canada}
\affiliation{Centre for Ultrahigh Bandwidth Devices for Optical Systems (CUDOS) \& MQ Photonics Research Centre, Department of Physics \& Astronomy, Macquarie University, Sydney, NSW 2109, Australia}
\author{Evan Meyer-Scott}
\affiliation{Institute for Quantum Computing, University of Waterloo, Waterloo, ON N2L 3G1, Canada}
\affiliation{Department of Physics and Astronomy, University of Waterloo, Waterloo, ON N2L 3G1, Canada}
\author{Amir~K. Khandani}
\affiliation{Department of Electrical and Computer Engineering, University of Waterloo, Waterloo, ON N2L 3G1, Canada}
\author{Norbert L{\"u}tkenhaus}
\affiliation{Institute for Quantum Computing, University of Waterloo, Waterloo, ON N2L 3G1, Canada}
\affiliation{Department of Physics and Astronomy, University of Waterloo, Waterloo, ON N2L 3G1, Canada}
\author{Thomas Jennewein}
\email{thomas.jennewein@uwaterloo.ca}
\affiliation{Institute for Quantum Computing, University of Waterloo, Waterloo, ON N2L 3G1, Canada}
\affiliation{Department of Physics and Astronomy, University of Waterloo, Waterloo, ON N2L 3G1, Canada}

\begin{abstract}

Quantum key distribution (QKD) has the potential to improve communications security by offering cryptographic keys whose security relies on the fundamental properties of quantum physics. The use of a trusted quantum receiver on an orbiting satellite is the most practical near-term solution to the challenge of achieving long-distance (global-scale) QKD, currently limited to a few hundred kilometers on the ground. This scenario presents unique challenges, such as high photon losses and restricted classical data transmission and processing power due to the limitations of a typical satellite platform. Here we demonstrate the feasibility of such a system by implementing a QKD protocol, with optical transmission and full post-processing, in the high-loss regime using minimized computing hardware at the receiver. Employing weak coherent pulses with decoy states, we demonstrate the production of secure key bits at up to \SI{56.5}{\dB} of photon loss. We further illustrate the feasibility of a satellite uplink by generating secure key while experimentally emulating the varying channel losses predicted for realistic low-Earth-orbit satellite passes at \SI{600}{\km} altitude. With a \SI{76}{\mega\Hz} source and including finite-size analysis, we extract 3374~bits of secure key from the best pass. We also illustrate the potential benefit of combining multiple passes together: while one suboptimal ``upper-quartile'' pass produces no finite-sized key with our source, the combination of three such passes allows us to extract 165~bits of secure key. Alternatively, we find that by increasing the signal rate to \SI{300}{\MHz} it would be possible to extract \num{21570}~bits of secure finite-sized key in just a single upper-quartile pass.

\end{abstract}

\maketitle

\section{Introduction}

Quantum key distribution (QKD) offers communications security without reliance on computational presumptions by taking advantage of fundamental properties of quantum mechanics~\cite{Bennett1984, Scarani2009}. Despite reaching maturity that supports commercial implementation~\cite{idQuantique, MagiQ}, QKD has yet to achieve widespread use, in large part owing to distance limitations, on the order of \SI{200}{\km}, inherent to lossy terrestrial transmissions~\cite{TNZHHTY07, Ursin2007, Schmitt07, Stucki_NJP250k, Liu_10, YHHJIJ12, korzh2014provably}. Quantum repeaters~\cite{Briegel1998} promise to overcome this shortcoming, but they require high-fidelity quantum memories which are still in the fundamental research stage~\cite{SAABDGHJKMNPRDRSSSTWWWWY10, Sangouard2011} and are not yet viable for real-world application. Alternatively, quantum links to orbiting satellites can be implemented using existing technologies~\cite{Gilbert2000, Nordholt02, Ursin2009, Bonato_NJP_09, Bourgoin2013}.

One near-term approach to satellite-based QKD is where a satellite, acting as a trusted node, performs two consecutive quantum key exchanges with two different ground stations. A combination of the two keys is then publicly revealed, allowing one ground station to extract the other's key, giving both locations a shared key in a way that no other party (except for the satellite) can surreptitiously intercept. This approach may be implemented with either uplink (photons sent from ground station transmitter to satellite receiver) or downlink (photons sent from satellite transmitter to ground station receiver). The feasibility of both of these has been extensively studied~\cite{BHLMNP00, Rarity_NJP_02, Bonato_NJP_09, Bourgoin2013}. While a downlink benefits from lower transmission losses, allowing higher key rates, an uplink offers the advantage of a simpler satellite design, easier pointing, reduced on-board data collection requirements, and source flexibility, which makes an uplink the preferred scenario for scientific study~\cite{MeyerScott2011, Bourgoin2013}.

A significant challenge in the uplink scenario is operating with the high photon loss experienced (\SIrange{40}{60}{\dB}). Previous work demonstrated that key extraction is possible, in principle, beyond \SI{50}{\dB} of loss in the infinite key limit~\cite{MeyerScott2011}. However, this work did not perform all of the steps necessary to implement the QKD protocol and produce a secure key. (Indeed, experimental QKD demonstrations routinely go no further than calculate the expected length of the secure key based on observed parameters.) Here we experimentally demonstrate key extraction at various transmission loss levels, up to \SI{56.5}{\dB}, while including all the QKD processing steps required to finally extract a secure key. We also examine the effect of finite statistics, and assess the time required to achieve near-asymptotic key rates in the high-loss environment. Further, we show the feasibility of ground--satellite QKD by experimentally recreating the varying losses of three realistic uplink satellite passes.

Our apparatus has the two parties involved in the high-loss QKD transmission operating independently, each party having separate event time-taggers, global positioning system (GPS) receivers, and classical processing mediated by a classical communication channel. Because we focus on a future satellite implantation, computational requirements are also a key aspect. The system we have developed attempts to reduce, as much as possible, these requirements at the receiver. We analyze the complexity of the classical processing functions, and demonstrate operation on low-power embedded hardware. We show that the requirements are feasible, making our overall design suitable for a satellite payload.

This paper is organized as follows. \Cref{sec:QKD} details the steps of the QKD protocol and the approaches we have taken to optimize it for a satellite uplink. \Cref{sec:Apparatus} describes the experimental apparatus we constructed to perform our demonstrations. \Cref{sec:Results} presents the results in two parts: \cref{sec:requirements} shows the results of our computational analysis, while \cref{sec:QKDresults} shows the results of our experimental QKD demonstration. We close with discussion and conclusion in \cref{sec:Conclusion}.

\section{Implementing QKD with limited resources}\label{sec:QKD}
\subsection{BB84 with decoy states}

The seminal QKD protocol, BB84~\cite{Bennett1984} encodes information in the polarization states of single photons. Ideally, at each time-step Alice randomly selects one of four polarizations in two bases---horizontal (H), vertical (V), diagonal (D), or anti-diagonal (A)---and sends a photon with this polarization to Bob. Bob randomly selects a basis, H/V or D/A, and measures the photon to obtain one of four outcomes. This procedure occurs for many time-steps, and after revealing the bases used, Alice and Bob ``sift'' their events, discarding those which have mismatched bases. By defining H and D to correspond to a bit value of 0, and V and A to correspond to a bit value of 1, Alice and Bob retain a common string of random bits---the \emph{sifted} key~\cite{Scarani2009}. 

Practical implementations have extra complications: photons are lost in transmission, imperfections in photon source and detection devices introduce errors (which must be corrected), and weak coherent pulse sources, which are often used in place of true single-photon sources, exhibit potentially insecure multi-photon emission events. Decoy-state protocols~\cite{Ma2005}, with error correction and privacy amplification as classical post-processing steps, have been developed to overcome these issues.

Theoretical QKD security proofs provide equations for the secure key rate based on experimentally measurable parameters such as the quantum bit error ratio (QBER), background noise counts, and decoy parameters. These allow us to make a statement about the security of the \emph{final} key after the quantum transmission is complete. Most importantly, these equations determine the amount of privacy amplification required to be able to claim $\epsilon$ security~\cite{Renner_phd, TLGR12}. Once the post-processing is complete and the final key deemed secure, it can then be used in a classical encryption protocol such as one-time pad.

\begin{figure}[tbp]
  \centering
  \includegraphics[width=\linewidth]{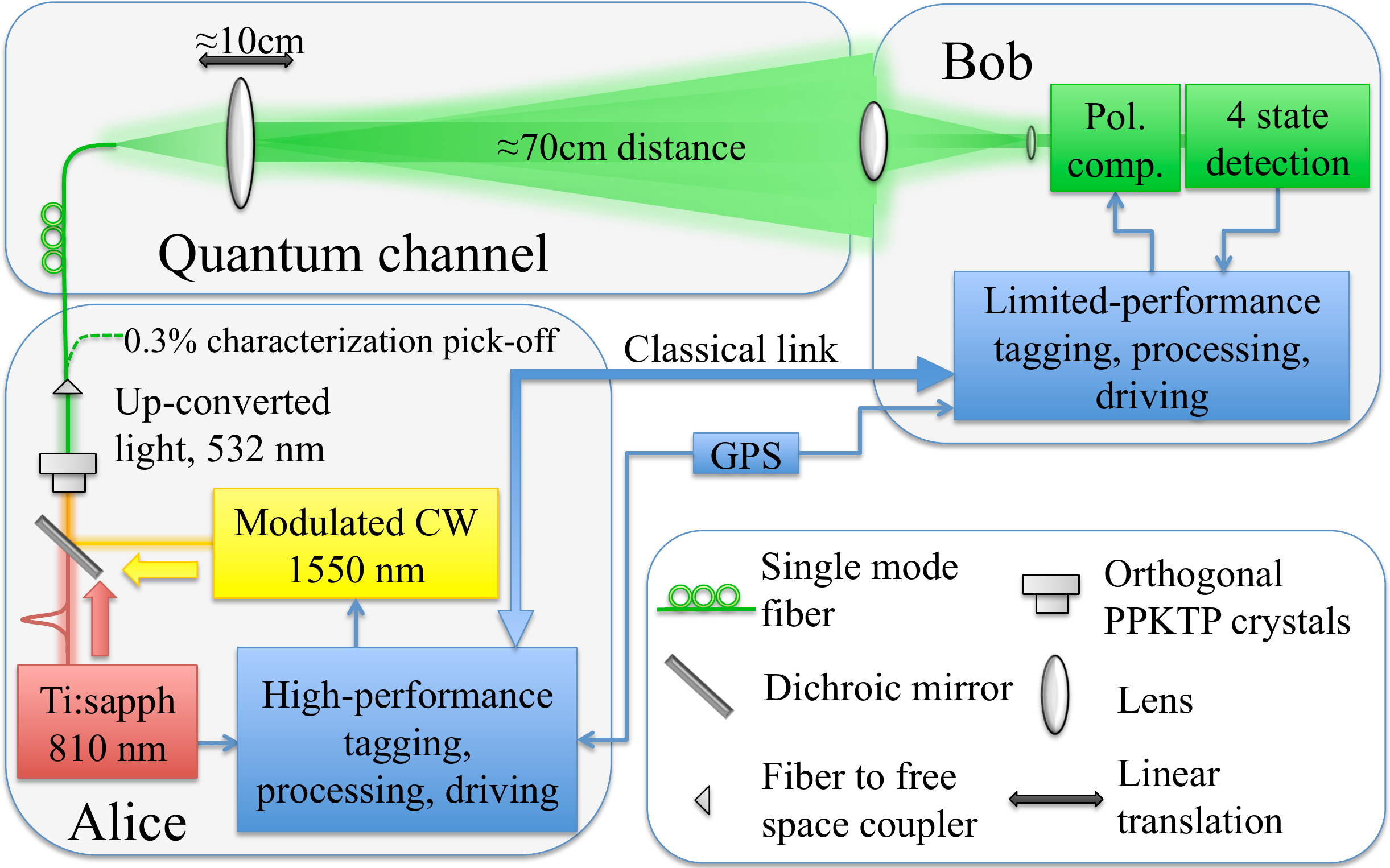}
  \caption{Schematic overview of our high-loss QKD apparatus. The source at Alice produces weak coherent pulses with wavelength \SI{532}{\nm} that possess both the short pulse length of the mode-locked \SI{810}{\nm} laser and the polarization state of the \SI{1550}{\nm} continuous wave laser. The quantum channel consists of a movable lens after the fixed output of an optical fiber to adjust the beam divergence over the free-space link. Computational performance of the tagging, processing and driving in Bob is limited to simulate the available resources of a satellite-based QKD receiver payload.}
  \label{fig.system_schematic}
\end{figure}

We implement the vacuum+weak decoy-state protocol~\cite{Ma2005}, in which Alice randomly emits signal states with average photon number $\mu$, or decoy states that are either vacuum or have an average photon number $\nu < \mu$. In our implementation (\cref{fig.system_schematic}), Alice employs polarization and intensity modulation~\cite{YMBHGMHJ13} to prepare a random sequence of BB84 polarization encodings which are 92\% signal and 8\% decoy states. Our average photon numbers are $\mu \approx 0.5$ and $\nu \approx 0.05$, which are near the optimal for this protocol. (Further details of the apparatus are given in \cref{sec:Apparatus}.)

The lower bound for the final asymptotic secure key rate per laser pulse is~\cite{Ma2005}
\begin{equation}\label{eq:SecureRate}
R_\infty = q K_\mu \left\{ - Q_\mu f_\text{EC} H_2(E_\mu)
+ Q_1^\text{L} \left[ 1 - H_2({E_1^\text{U}}) \right] \right\}.
\end{equation}
Here, $q$ is a basis reconciliation factor ($\sfrac{1}{2}$ for BB84), $K_\mu$ is the fraction of pulses that are signal states, $f_\text{EC}$ is the efficiency parameter of the error correction algorithm, $H_2$ is the binary entropy function, $Q_{\mu/\nu}$ is the gain for signal/decoy states (the ratio of number of photons detected by Bob to number of pulses sent by Alice), $E_{\mu/\nu}$ is the QBER for signal/decoy states (ascertained in the error correction process), and $Q_1^\text{L}$ and $E_1^\text{U}$ are the lower bound of the gain and the upper bound of the QBER, respectively, for single-photon pulses.

The single-photon gain lower bound $Q_1^\text{L}$ is calculated as
\begin{equation}\label{eq:Q1L}
Q_1^\text{L}  = \frac{\mu^2 e^{-\mu}}{\mu\nu - \nu^2}\left(Q_\nu e^\nu - Q_\mu e^\mu \frac{\nu^2}{\mu^2}
- \frac{\mu^2 - \nu^2}{\mu^2 }Y_0\right)
\end{equation}
where $Y_0$ is the vacuum yield, determined by the cumulative probability of detector dark counts and background noise within the coincidence window. The single-photon QBER upper bound $E_1^\text{U}$ is calculated as~\cite{Ma2005, CaiScarani}
\begin{align}
E_1^{\text{U},\mu}  & = \frac{E_\mu Q_\mu}{Q_1^\text{L}} - \frac{E_0 Y_0}{Q_1^\text{L} e^\mu} \\
E_1^{\text{U},\nu}  & = \frac{E_\nu Q_\nu e^\nu - E_0 Y_0}{\nu Q_1^\text{L} }\mu e^{-\mu} \\
E_1^\text{U}       & = \min\left\{E_1^{\text{U},\mu}, E_1^{\text{U},\nu}\right\} \label{eq:e1U}
\end{align}
where $E_0$ is the vacuum error rate ($\sfrac{1}{2}$ in a perfect apparatus).

In the present work, the parameters in equations (\ref{eq:SecureRate})--(\ref{eq:e1U}) are determined from experimental data to obtain the asymptotic lower bound for the secret key rate per laser pulse, $R_\infty$. To obtain the secure key rate in bits per second, $R_\infty$ is multiplied by the pulse rate of the WCP source. 

Proper generation of a secure key needs to incorporate the effects of statistical fluctuations due to finite-sized experimental data~\cite{LPDFSDYPS13}. To account for this we use the common heuristic of adding or subtracting $10\sigma$ variation from the experimental parameters in such a way as to minimize the key rate~\cite{Sun2009}. (A recently proposed method may allow to account for statistical fluctuation in a more rigorous fashion~\cite{curty2014finite}.) Finite-size security effects are captured~\cite{ScarRenner08} by the security parameter $\Delta$, resulting in a key rate lower bound
\begin{align}\label{eq:FiniteSizeRate}
R &= q K_\mu \Big\{ - Q_\mu f_\text{EC} H_2(E_\mu) \nonumber\\
&\quad\quad + Q_1^\text{L} \left[ 1 - H_2({E_1^\text{U}}) \right] - Q_\mu \Delta / N_\mu \Big\}
\end{align}
where $\Delta = 7\sqrt{N_\mu \log_2[2/(\bar\epsilon - \bar\epsilon')]} - 2\log_2 [2(\epsilon - \epsilon_\text{EC} - \bar\epsilon)]$, $\epsilon_\text{EC}$ is the error correction silent failure probability (we use $10^{-10}$), $N_\mu$ is the raw key size, and $\bar\epsilon$ and $\bar\epsilon'$ are numerically optimized for $R$, constrained by $\epsilon - \epsilon_\text{EC} > \bar\epsilon > \bar\epsilon' \geq 0$.

\subsection{Distribution of post-processing tasks}

The design of our classical post-processing software follows the principle that Alice should perform as many of the computationally intensive tasks as possible, as the ground station can be made rich in computing resources, compared to the limited capacity of a satellite payload. In our system, Alice, being the source of the optical signal over the high-loss link, is responsible for high-rate data readout. She also performs timing analysis (to match Bob's classically transmitted time-tagged photon detection events to her time-tagged source events) and basis sifting, afterwards sending simplified coincidence information back to Bob. We also choose a one-way error correction algorithm based on low-density parity-check codes~\cite{MN97}, in which Alice performs the computationally expensive decoding algorithm while Bob only runs a linear algorithm to compute his syndromes (see \cref{sec:LDPC}). This scheme has the additional advantage of having low classical communication overhead. Finally, both parties perform a Toeplitz-matrix-based~\cite{Krawczyk94} privacy amplification routine suitable for low-power hardware implementation (see \cref{sec:PA}). 

We separate Bob's software into two components: a driving control environment, and an embedded processing component. The driving control component is responsible for all platform-dependent tasks, e.g.\ loading time-tagger operating system drivers, configuring time-taggers, reading out time-tags and displaying live statistics. To be suitable for implantation into a yet-to-be-designed satellite, Bob's embedded processing component is implemented in a platform-agnostic way using a portable low-level language (C). It is executed as a separate process on an x86-64 desktop computer, or on a low-power ARM development board, and performs the bulk of Bob's necessary processing tasks.

Because Bob's embedded component runs in a standalone process, its usage of computing resources can be accurately monitored. Moreover, the driving control component records the bandwidth used for classical communication. This design allows us to make an accurate assessment of the classical post-processing requirements and guide our analysis of the computing requirements of Bob's part of the QKD protocol.

\subsection{Time synchronization and basis sifting}

Several practical issues complicate the process of determining which time-tagged photon detections correspond to particular source events, including initial clock synchronization, drift over time, and variation in photon time-of-flight. For our apparatus, an extra complication is that our data acquisition hardware is not capable of operating at the high pulse frequency of the laser source (see \cref{sec:Apparatus}).

To reduce the heavy load on the time-tagger at the source, only a subset of the laser's output pulses are time-tagged. With Alice utilizing a predefined (known only to Alice) randomized sequence of pulse states, we assume that the laser's period is stable on the order of \si{\micro\s}, and interpolate to reconstruct the transmitted states for timing analysis. The predefined sequence is not a requirement of the time-tagger but is necessary due to limitations of the modulation electronics at the source, which require a preloaded sequence.

To reduce clock drift, we align the time-tagging units' internal clocks to a \SI{10}{\MHz} time-base signal provided by a GPS receiver at each site. Initial synchronization is achieved with the one pulse per second (\SI{1}{PPS}) signal provided by the GPS receivers. Position data is supplied with these signals, which can be used in conjunction with time data to estimate the distance between Alice and Bob, and hence, the time-of-flight of the photons between the source and the receiver.

In our system, photon detections are tagged with a resolution of \SI{156.25}{\ps}, but the \SI{1}{PPS} signals are accurate to ${\approx}\SI{100}{\ns}$. Additional analysis is required to identify corresponding emission and detection events to within a desired coincidence window of about \SIrange{0.1}{1.3}{\ns}. The algorithm to achieve this synchronization utilizes the timing information from Bob's time-tags, Alice's transmitted photon states, Alice's and Bob's GPS timing and position data, as well as a small subset (${\approx}\SI{5}{\percent}$) of Bob's measured outcomes. Alice employs a histogram-based optimizing coincidence search within a predefined time span (about \SI{100}{\ns}). The subset of Bob's revealed measured outcomes (which are discarded from the final key) are also used to estimate the QBER to commence the error correction stage of the QKD protocol.

Once coincident events have been identified and noncoincident events removed, Alice performs basis sifting of her raw key to produce her sifted key and transmits a list of indices to Bob, which he utilizes to equivalently sift (for both time and basis, simultaneously) his photon detection events.

\subsection{Error correction}\label{sec:LDPC}

Low-density parity-check (LDPC) codes are highly suitable for satellite-based QKD due to the low communication overhead required and the inherent asymmetry in the computational complexity at each site. First, Alice prepares an $M \times N$ irregular~\cite{Hu_peg} parity-check matrix, where $N$ is the sifted key block size, and $M$ is based on the QBER estimate obtained during timing analysis. We use progressive-edge growth software~\cite{Hu_peg} (modified from~\cite{PEG_SW}), employing known optimal degree distribution profiles~\cite{El_2009, Mateo_phd}, to generate the parity-check matrix. Alice then transmits the matrix in a compact form to Bob over a classical channel. For each $N$-bit block of his sifted key, Bob runs an efficient linear algorithm to compute a syndrome using this matrix, and transmits it to Alice.

Alice then attempts to reconcile her sifted key assuming that Bob's sifted key is ``correct'' (it remains unchanged throughout this process). For each block of sifted key, Alice's goal is to resolve Bob's key vector $\mathbf{x}$, based on her key vector $\mathbf{y}$, Bob's syndrome $\mathbf{s}$, the parity-check matrix, and the estimated QBER. To accomplish this task, Alice employs \emph{belief propagation}, an iterative message passing decoding algorithm, also known as the sum-product algorithm~\cite{Pea04, Lucio-Martinez_NJP_09}. Our sum-product LDPC decoder is written in C\# and is based on that found in~\cite{Chan_masc}. Upon success, Alice and Bob both possess the $N$-bit error-corrected key block $\mathbf{k}_\text{EC} = \mathbf{x}$ and obtain the exact QBER for the quantum transmission, $E_\mu$.

By Shannon's channel coding theorem~\cite{Shannon1948} applied to the binary symmetric channel~\cite{CoverIT}, we can deduce a closed-form estimate of the appropriate size of the LDPC matrix based on the (estimated, denoted by a tilde) QBER~\cite{El_2011}: $M = N f_\text{EC} H_2(\tilde{E}_\mu)$. The decoding step may yet terminate unsuccessfully with a given matrix and key block---the probability of this decreases as $M$ (and thus $f_\text{EC}$) is increased. In the case of such a termination, we may either discard the key block or retry the algorithm with an augmented matrix containing all the rows of the previous matrix, similar to the ``nested'' LDPC codes proposed in~\cite{Rav_NestedLDPC}. In a satellite mission, the choice can be based on the availability of the classical communication channel. Our implementation exhibits efficiencies ($f_\text{EC}$) ranging from 1.1 to 1.5.

The silent failure probability of the belief propagation procedure---i.e., the probability that the process terminates successfully but there remains one or more uncorrected bits---is not well characterized in existing literature. While we have not observed any silent failures during our testing, we cannot be certain that $\epsilon_\text{EC} = 10^{-10}$ is achieved. To ensure such certainty, one could calculate, reveal, and compare a fingerprint hash of $\mathbf{x}$ and $\mathbf{k}_\text{EC}$. (Such an approach using 128-bit MD5 sums~\footnote{Although MD5 is not cryptographically secure as methods to modify a file while preserving its hash value are known, this does not assist Eve's effort to reconstruct the \emph{same} key as Alice and Bob.}, for example, yields a collision probability, and thus a silent failure probability, of order $2^{-128} \approx 3\times10^{-39}$.) To account for the revealed bits, the final key length would need to be reduced by the same (constant) number of bits. Because the necessity of these extra steps and the specific method of implementation are unclear, we do not perform these steps here.

\subsection{Privacy amplification}\label{sec:PA}

The error-corrected key block $\mathbf{k}_\text{EC}$ is only partially secure, as some information may have leaked to an eavesdropper (Eve)---we attribute the observed QBER to Eve's interaction with the signal, and all parity information communicated during error correction is known to Eve as it was transmitted over a public channel. Privacy amplification is employed to create a new, final, key $\mathbf{k}_\text{F}$ on which Eve no longer holds more than negligible amount of information. The procedure consists of applying a \emph{two-universal hash function}~\cite{Scarani2009, Tsurumaru_11_PA} to $\mathbf{k}_\text{EC}$ to produce a provably secure key block $\mathbf{k}_\text{F}$ of length $L < N$ (recall $N$ is the sifted key block size). $L$ is obtained by multiplying $R$ of \cref{eq:FiniteSizeRate} by the number of pulses sent. For mitigating the nonlinear length reduction due to finite-size effects, $N$ should be kept above a certain value, typically ${\sim}10^5$, as finite-size effects heavily impact keys with lower $N$. This value is taken into consideration when selecting a hash function.

Privacy amplification is a symmetric operation which needs to be performed by both Alice and Bob. The choice of hash function dictates the computational complexity of the process and the amount of classical communication required. In our implementation, the privacy amplification procedure loosely follows the methodology outlined in~\cite{Tsurumaru_11_PA}---however, we have made some alterations to their model and developed a different matrix multiplication procedure suitable for efficient implementation in hardware. Briefly, we employ the Toeplitz matrix~\cite{Gray_2005} construction implemented using a shift register.

A Toeplitz matrix has constant descending left-to-right diagonal elements. An $L \times N$ Toeplitz matrix can be written as
\begin{equation}
 T_\mathbf{r} = \begin{bmatrix}
      r_L       & r_{L+1} & \cdots      &     &       &   &  \cdots & r_{N+L-1} \\
      r_{L-1} & r_L        & r_{L+1} & \cdots     & &    &  \cdots & r_{N+L-2}   \\
       \vdots  &   & \ddots    &  &       &  	&        & \vdots \\
      r_2       &   \cdots  & r_{L-1}   &	r_L		& r_{L+1}     & \cdots &  \cdots  & r_{N+1}\\
      r_1       & r_2        & \cdots     &	r_{L-1}		& r_{L}         & r_{L+1}	& \cdots            & r_{N}  
     \end{bmatrix}.
\end{equation}
A Toeplitz matrix is a two-universal hash function~\cite{Krawczyk94}. Note that a Toeplitz matrix $T_\mathbf{r}$ is completely defined by the $(N+L-1)$-bit vector $\mathbf{r} = (r_1, r_2, \dots, r_{N+L-1})$, thus its storage and transmission requirements are considerably reduced. Further, an $L \times N$ matrix of the form $U_\mathbf{r} = (I_{L} | T_\mathbf{r})$, i.e.\ a concatenation of an $L$-dimensional identity matrix $I_{L}$ and an $L \times (N-L)$ Toeplitz matrix $T_\mathbf{r}$, is also a two-universal hash function as we require, but requires only $N-1$ bits to define~\cite{Golub_matrix, Hayashi_11_PA}.

Following error correction, Alice generates such a matrix by constructing a random binary string $\mathbf{r} = (r_1, r_2, \dots, r_{N-1})$ of length $N-1$, and then transmits $\mathbf{r}$ to Bob over the classical channel. Alice and Bob then use $\mathbf{r}$ and a shift register to apply the hash matrix $U_\mathbf{r}$, computing the final secure key, $\mathbf{k}_\text{F} = U_\mathbf{r} \mathbf{k}_\text{EC}$.

In our implementation, the identity portion of each row of $U_\mathbf{r}$ uses no space and can be accounted for with a simple logical AND operation. We represent $T_\mathbf{r}$ as an $(N-L)$-bit logical shift register. Initially, the shift register contains the last $N-L$ bits of $\mathbf{r}$, $(r_{L}, r_{L+1}, \dots, r_{N-1})$. The remaining bits from $\mathbf{r}$ are used as input for the shift register. In this way, we conserve memory by never needing to store full matrices.

The logical shift register is broken up into multiple 32-bit blocks, each of which is designed to fit inside a register on a processing unit. The register size of 32 bits is chosen for the support of multiple platforms, including our low-power ARM test board. 64-bit platforms are also available, and with single instruction, multiple data (SIMD) extensions, commonplace in contemporary desktop processors, the register could be 256 bits or larger.

After privacy amplification, Alice and Bob are left with a secure key of $L$ bits which can then be used to encrypt data transmitted on a classical channel through, e.g., one-time pad. We assume here that channel authentication is performed separately, possibly using some of the secure key (reducing the length available for encryption).

\section{Apparatus}\label{sec:Apparatus}

Our QKD system, shown in \cref{fig.system_schematic}, consists of a weak coherent pulse (WCP) source, a variable-loss free-space channel, and a compact four-outcome (two passively chosen measurement bases) quantum receiver. The source utilizes up-conversion (sum frequency generation) from two orthogonally oriented type-I periodically poled potassium titanyl phosphate (PPKTP) crystals to produce photon pulses at \SI{532}{\nm} wavelength from a mode-locked Ti:sapphire laser at \SI{810}{\nm}, operating at a rate of \SI{76}{\MHz}, and a continuous-wave laser at \SI{1550}{\nm}.

Diagonally polarized \SI{810}{\nm} laser pulses are combined with polarization- and intensity-modulated \SI{1550}{\nm} laser light (controlled by efficient telecom waveguide modulators~\cite{YMBHGMHJ13}) to generate \SI{532}{\nm} pulses possessing the short pulse width and high repetition rate of the \SI{810}{\nm} laser, as well as the intensity and polarization of the \SI{1550}{\nm} light. Phase randomization between pulses, necessary to ensure security, is provided by the short coherence time of the \SI{1550}{\nm} laser (less than the pulsing period of the \SI{810}{\nm} laser, but much more than the pulse duration). Birefringent wedges precompensate the \SI{810}{\nm} light for temporal walk-off in the PPKTP crystals.

The photon pulses produced are coupled into single-mode fiber. A fiber splitter sends ${\approx}\SI{0.3}{\percent}$ of photons to a thick-silicon avalanche photodiode (Excelitas SPCM-AQ4C) to measure the average photon number per pulse. The remaining photons are sent to a free-space quantum channel consisting of a bare fiber output followed by a 3-inch-diameter lens on a longitudinal translation stage. The loss is adjusted by varying the position of the lens, changing the amount of light directed into the receiver by making the beam more or less divergent.

\begin{figure}[tbp]
  \centering
  \includegraphics[width=\linewidth]{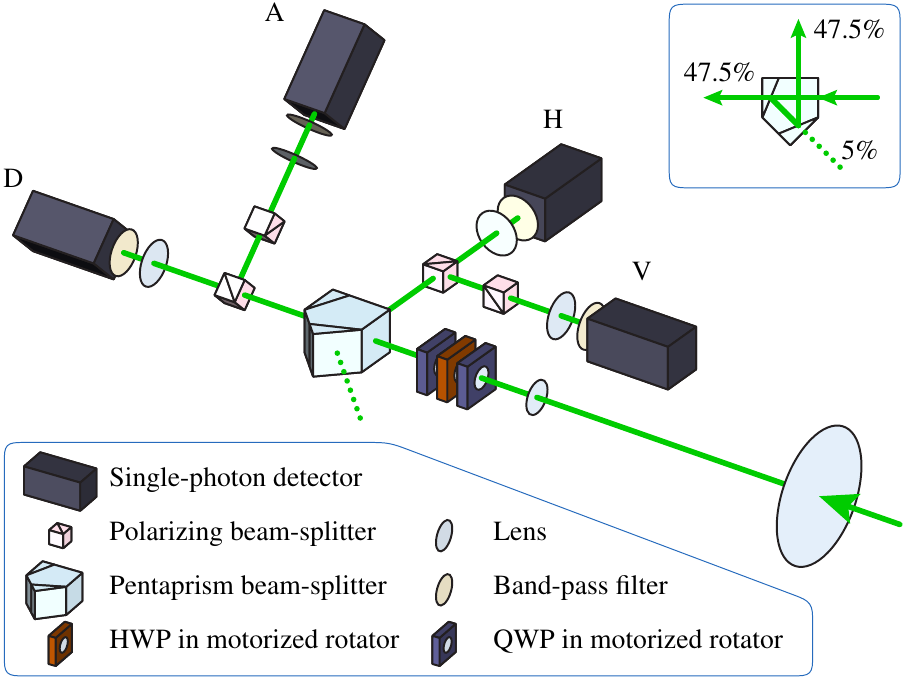}
  \caption{High-loss QKD receiver. Photons are captured by the telescope and pass through motorized-rotating half- and quarter-wave plates, correcting unwanted polarization rotations. The pentaprism beam-splitter provides a passive basis choice between rectilinear and diagonal polarization bases. A polarizing beam-splitter (PBS) in each basis arm discriminates H/V and D/A polarized photons, the latter by physically orienting the PBS \SI{45}{\degree} around the beam path. An extra PBS in each reflected arm reduces erroneous counts from device imperfections. Lenses focus the beams onto single-photon detectors, while band-pass filters reduce background light.}
  \label{fig.receiver_schematic}
\end{figure}

The quantum receiver, \cref{fig.receiver_schematic}, is built using Thorlabs' cage system. The receiving telescope consists of a \SI{5}{\cm} diameter, \SI{25}{\cm} focal length collection lens followed by a \SI{6.5}{\mm} diameter, \SI{11}{\mm} focal length collimating lens. Passive measurement basis choice is implemented by coupling polarization discrimination apparatuses to two orthogonal outputs of a pentaprism beam-splitter, each of which transmits approximately \SI{47.5}{\percent} of the injected power. (A pentaprism was chosen for potential application in future experiments---a 50:50 beam-splitter would also suffice as we do not here use the pentaprism's third output.)

Polarization analysis is done by a \SI{5}{\mm} cubic polarizing beam-splitter (PBS) in each arm, directing photons to one of four detector assemblies. Measurement in the diagonal basis is obtained by physically orienting (to \SI{45}{\degree}) the PBS and detectors around the beam path, relative to the other PBS (which defines the rectilinear basis). Following each analysis PBS, a second PBS, oriented at \SI{90}{\degree}, is used to suppress erroneous optical signal in the reflected path (owing to device imperfection).

Each detector assembly contains a spatial-filtering shield and a \SI{2}{\nm} band-pass filter to suppress background noise. Photons are focused by \SI{6}{\cm} focal length lenses and detected by thin-Si avalanche photodiodes from Micro Photon Devices, which feature good detection efficiency (${\approx}\SI{50}{\percent}$), low dark counts (${\approx}\SI{20}{cps}$) and low jitter (${\leq}\SI{50}{\ps}$). Temporal filtering with a narrow (${\sim}\SI{1}{\ns}$) time window allows us to accept signal photons while rejecting remaining background and dark counts with high fidelity. The background yield $Y_0$ is estimated by counting photon detections between pulses. Though this approach is known to be insecure, it suffices for our proof-of-concept demonstration.

Data are acquired by two time-tagger units, and processed by two x86-64 computers (and, when testing algorithmic performance, an ARM board) on a local-area network (LAN). Each time-tagger unit is connected to \SI{10}{\MHz} and \SI{1}{PPS} signals coming from a GPS receiver. The signals from the source are connected to Alice's time-tagger and the four outputs of Bob's detectors are connected to Bob's time-tagger.

Our receiver also includes an arbitrary polarization rotation assembly, consisting of quarter-wave plates (QWPs) before and after a half-wave plate (HWP) mounted in motorized rotation stages, allowing the compensation of any unitary polarization change in the channel. We have developed an automated polarization alignment protocol which characterizes the effect of the channel on the known QKD polarization states, measuring the quantum signal directly, sufficient to then determine an optimal compensation implemented by the arbitrary polarization rotation assembly.

\section{Results}\label{sec:Results}

\subsection{Post-processing resource requirements}\label{sec:requirements}

For a satellite uplink scenario, optical signals are sent from the ground when the satellite orbits over an optical ground station, while classical communication is performed---possibly at a later time---when the satellite orbits over one or more radio frequency (RF) ground stations. Hence, the satellite system must store all time-tags accumulated during the optical station flyover, performing all classical steps of the QKD protocol during an RF station flyover when a classical communication link is present. Specifically, Bob must store: time-tags, measurement bases, photon detections (bit values), and data defining the error correction LDPC matrix and privacy amplification Toeplitz matrix.

Our time-tagging hardware produces 64-bit time-tags. To save on the limited memory and classical communications bandwidth available to a satellite, it is possible to reduce that number significantly, at the expense of additional computation steps. One simple scheme is to store the full time-tag only at the beginning of every second of data collection, together with additional information provided by the GPS receiver (which outputs a data packet every second). In this way, the space required to store the time-tag and measurement outcome information is 40~bits.

To further save memory and classical communications bandwidth, the sparse LDPC matrix for error correction can be efficiently transmitted and stored as an adjacency list where only the indices of each non-zero element in each row are recorded. If the decoding step fails, we must then retry with a larger matrix (or discard the block), implying an increase of $f_\text{EC}$, i.e., worse efficiency.

The embedded processing component of the satellite-side software is tested on an inexpensive (${\approx}\$150$), low-power (\SI{2}{\W}) Freescale i.MX53 QSB single-board computer featuring a \SI{1}{\GHz} single-core ARM processor and \SI{1}{GiB} of RAM.
The measured performance, \cref{tab:cpustats}, illustrates successful operation within reasonable resource constraints.
We have found that in our system the limiting factor for Bob is the privacy amplification step, which requires a relatively long processing time. For all other processes, the limiting factor is not Bob's computational power, but rather Alice's, and the \SI{100}{\mega\bit} Ethernet link between them. A future implementation could have a far more powerful computer at Alice than what we have used for our demonstration.

\begin{table}[t]
  \caption{Measured performance of the satellite-side QKD process running on a Freescale i.MX53 embedded ARM board processing 300~seconds of QKD data (\SI{28.8}{\dB} loss data with rate-limiting applied; see text). Here, privacy amplification is applied without incorporating finite-size effects which reduce secure key length, giving us upper bounds on resource usage. As expected, processing time scales quadratically with the photon detection rate---a least-squares quadratic fit gives a coefficient of determination $R^2 = 0.9992$.}  
\centering
    \begin{tabular}{S[table-format=5]S[table-format=5]S[table-format=1.2]S[table-format=4.1]S[table-format=2.2]}
    \hline\hline
{\textbf{Detection}} & {\textbf{Sifted key}} & {\textbf{QBER}} & {\textbf{Processing}} 	& {\textbf{RAM used}} \\
{\textbf{rate [Hz]}} & {\textbf{rate [Hz]}} & {\textbf{[\%]}} & {\textbf{time [s]}} & {\textbf{[Mbyte]}} \\
    \hline
500 & 229 & 3.43 & 0.5 & 11.19 \\
1000 & 457 & 3.44 & 1.1 & 20.42 \\
5000 & 2326 & 3.53 & 14.5 & 70.09 \\
10000 & 4647 & 3.57 &  56.3 & 71.70 \\
20000 & 9286 & 3.55 & 772.1 & 75.47 \\
30000 & 13924 & 3.54 & 2013.1 & 79.38 \\
41887 & 19428 & 3.54 & 3969.4 & 84.52  \\
    \hline\hline
    \end{tabular}
  \label{tab:cpustats}
\end{table}

For computational requirements analysis, we collect experimental data for 300~seconds at a receiver detection rate of about \SI{42}{\kilo\Hz}. Each one-second chunk of this data is then truncated to produce various effective detection rates. \Cref{tab:cpustats} shows detailed memory and CPU usage for the embedded processing component. Privacy amplification complexity is asymptotically quadratic in the block size $N$ due to the matrix multiplication process, while all other post-processing steps behave linearly. Hence, we expect the processing time of the QKD post-processing to overall scale quadratically with the detections, as is observed. Note that we do not expect the number of detections to exceed \num{1e7} over a single satellite pass using feasible quantum sources~\cite{Bourgoin2013}.

\subsection{Experimental secure key extraction}\label{sec:QKDresults}

\begin{figure}[tbp]
  \centering
 \includegraphics{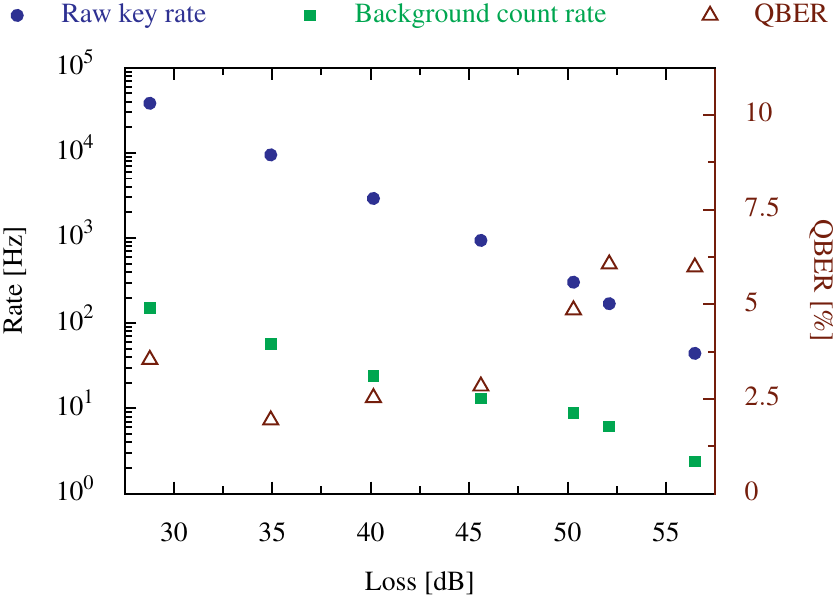}
  \caption{Measured raw key rate, background detection rate and QBER obtained in different loss regimes, with a source pulsing at \SI{76}{\MHz}. The raw key and background rates include only detections that fall within the temporal filter window. The background rate (the product of the vacuum yield $Y_0$ and the pulsed laser frequency) is determined by measuring the counts received between laser pulses. At lower loss, the background term is dominated by light from the \SI{1550}{\nm} laser and some continuous wave component remaining in the pulsed \SI{810}{\nm} laser. Variations in QBER between runs are mainly due to laboratory temperature fluctuations which affected the birefringence of the optical fiber and the performance of the \SI{1550}{\nm} modulators. Loss includes both channel loss (variable) and receiver efficiency (fixed \SI{4.5}{\dB}).}
  \label{fig.QBER_raw_rate}
\end{figure}

We perform the experimental demonstration for losses ranging from 28.8 to almost \SI{60}{\dB}, determined from the photon detection rate (corrected for background) with respect to the transmitted optical power. The loss therefore includes both channel loss (variable) and receiver efficiency (fixed \SI{1.5}{\dB} for receiver optics and \SI{3}{\dB} for detector efficiency). The temporal filter window width is adjusted to improve the secure key rate for each value of loss, and ranges from \SI{1.3}{\ns} at low loss to \SI{0.1}{\ns} at high loss. The measured QBER of signal states ranges from \SIrange{1.94}{6.06}{\%}, with raw key rate (total detections within the temporal filter window, per second) ranging from \SI{38211}{\Hz} at \SI{28.8}{\dB} to \SI{44.2}{\Hz} at \SI{56.5}{\dB}, while the background detection rate ranges from \SIrange{151}{2.38}{\Hz} (see \cref{fig.QBER_raw_rate}).

\begin{figure}[tbp]
  \centering
  \includegraphics{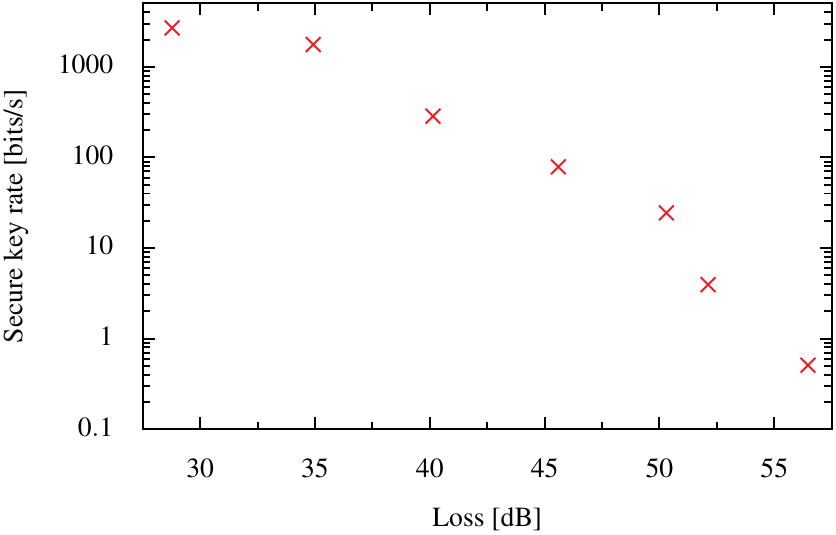}
  \caption{Secure key rate (lower bound) in the infinite limit for data measured in different loss regimes. The secure key rate tends to decrease as the loss increases, with some fluctuation about the trend due to variations in the source tuning and channel parameters throughout the data collection campaign. Loss includes both channel loss and receiver efficiency.}
  \label{fig.fixed_loss}
\end{figure}

Our experimental results incorporate the full error correction and privacy amplification post-processing. To limit computational time we artificially restricted the error correction block size to \num{600000} (with the sifted key split into the necessary number of blocks). Privacy amplification was implemented over the full sifted-key length of error-corrected key bits in order to minimize finite size effects. We achieve error correction efficiencies between 1.12 and 1.50 (with better efficiencies at higher QBER, as predicted by~\cite{TMPE14}) and privacy amplification to $\epsilon = 10^{-9}$ security. The extracted secure key rate is shown in \cref{fig.fixed_loss} for asymptotic extrapolations to the infinite limit of key length (\cref{eq:SecureRate}). At the highest loss, \SI{56.5}{\dB}, our system is able to extract \SI{0.5}{\bit/\s} of secure key in the asymptotic limit. This is comparable to the result of a previous high-loss demonstration~\cite{MeyerScott2011} which reached \SI{2}{\bit/\s} at \SI{57}{\dB} (the achievable rate there being inferred without implementing the complete QKD protocol). We note that the key rate can be readily improved by employing a faster source---QKD WCP sources have been demonstrated in the \si{\GHz} range~\cite{TDLFY06, TNZHHTY07}, more than an order of magnitude above our \SI{76}{\MHz} source rate.

\begin{figure}[tbp]
  \centering
  \includegraphics{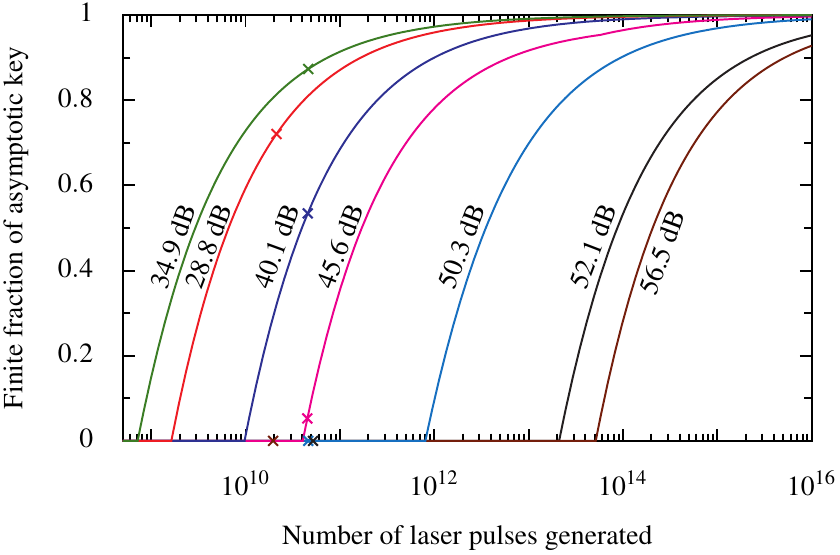}
  \caption{Finite-sized secure key rate as a fraction of the corresponding asymptotic secure key rate, given the total number of laser pulses transmitted. Curves are shown for experimental parameters corresponding to each of the loss conditions demonstrated, as indicated by labels beside each curve. Crosses indicate the value that was reached in the experiment. Lowest losses only require around $10^{10}$ pulses to exceed \SI{80}{\percent} of the asymptotic key rate. For the highest losses to reach this amount, significantly more pulses, $10^{14}$ to $10^{15}$, must be transmitted due to the reduced signal-to-noise. Note these curves indicate secure key rates relative to those that could be achieved in the asymptotic limit, not absolute rates.
}
  \label{fig.finite_time}
\end{figure}

Given that the apparatus remains sufficiently stable, the particular finite duration over which data are collected in this experiment is arbitrary. For our results, each data run lasts \SIrange{5}{10}{min}. With such times, when incorporating finite-size statistics (\cref{eq:FiniteSizeRate}) we find positive secure key rates for points up to \SI{45.6}{\dB}. For higher losses, there is insufficient statistics to produce nonzero key under the condition of $10\sigma$ worst-case variation. As the detriment of finite-size effects is due to the limited number of photon counts, a faster source can also mitigate this.

Based on the measured experimental parameters, we can extrapolate the raw key rates and determine the achievable finite-size secure key rate if the apparatus was run for longer times or multiple runs under the same conditions were concatenated. \Cref{fig.finite_time} shows the ratio of the asymptotic key rate that can be achieved by the finite secure key ($R/R_\infty$), for a given total number of pulses transmitted, for each experimental loss condition we examine. For the lowest losses, only about $10^{10}$ pulses is necessary to reach over \SI{80}{\percent} of the asymptotic key rate, equating to a few minutes of collection time with our \SI{76}{\MHz} source. More time is required for higher losses: several weeks of continuous collection at \SI{76}{\MHz}, for the highest losses. Interestingly, we find that the \SI{34.9}{\dB} data produces nonzero secure key sooner than the \SI{28.8}{\dB} data---this is owing to the relatively high QBER at \SI{28.8}{\dB}. Our results consistently show a significantly higher decoy QBER compared to the signal QBER. This was caused by the intensity modulator which was found to produce a slight polarization shift that is dependent on the applied modulation, causing the two different intensity levels to have slightly different polarizations before being polarization modulated, leading to a difference in the optimal alignment for the two intensity levels. This polarization shift could be corrected by the addition of a polarizer after the intensity modulator, eliminating the polarization difference between the two states. Although this difference does not invalidate our proof-of-concept demonstration, removing it is crucial in a secure implementation as it leads to distinguishability between signal and decoy states which could be used by an eavesdropper to gain information. Removing this difference may also improve final key rates as it would reduce decoy QBER without affecting signal QBER. Our system alignment is optimized for the signal QBER.

\begin{figure}[tbp]
  \centering
  \includegraphics{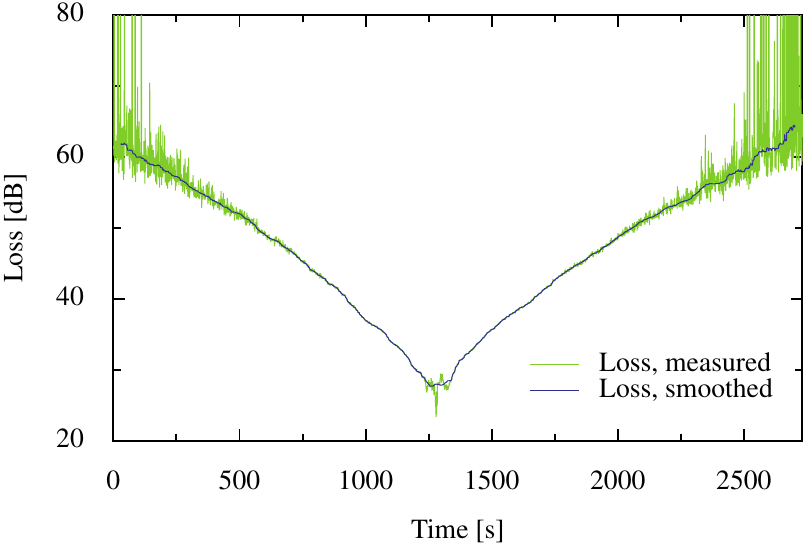}
  \caption{Experimentally measured loss over the \SI{45}{min} data collection used to simulate the varying loss of a satellite pass. The data are smoothed by taking the median value of a \SI{29}{\s} moving window. These smoothed values are used to select experimental data that tracks the theoretical loss of a satellite pass while maintaining the natural statistical fluctuations.}
  \label{fig.varying_loss}
\end{figure}

The results presented here are comparable to the regime of a satellite uplink, where the usable part of a pass is expected to vary typically between \SIrange{40}{55}{\dB} loss~\cite{Bourgoin2013}, and help support the conclusion that our approach is suitable for eventual satellite implantation (though a faster source may be advised). To more closely examine the feasibility of our approach in the regime of a satellite uplink for QKD, we simulate several satellite passes by varying the position of the quantum-channel lens (thus varying the loss) during an experimental run. We do this once, the total run lasting approximately 45~minutes, with the loss changing smoothly from $\approx$\SI{63}{\dB} to a minimum of $\approx$\SI{28}{\dB} after about 21~minutes, and then back to $\approx$\SI{62}{\dB} over the remaining time. The data accumulated are segmented into \SI{1}{\s} blocks, with the measured loss for each second over the duration of the experiment shown in \cref{fig.varying_loss}.

We redistribute select \SI{1}{\s} blocks of raw key data in such a way that we obtain data sets that reproduce the statistics expected for real satellite uplink orbits~\cite{Bourgoin2013}. The passes considered are the best, upper-quartile and median passes (in terms of contact time) over a hypothetical ground station located at \SI{45}{\degree} latitude of a year-long \SI{600}{\km} circular Sun-synchronous low Earth orbit. The predicted losses are based on uplink at a wavelength of \SI{785}{\nm}, with a receiver diameter of \SI{30}{\cm}, a \SI{2}{\micro rad} pointing error and a rural sea-level atmosphere. The differences with our system (which has \SI{532}{\nm} wavelength and \SI{5}{\cm} receiver diameter) are necessary to mitigate the increased geometric losses over the long distance link of a satellite (requiring larger receiver diameter) and the effect of atmospheric turbulence and transmission (reduced at \SI{785}{\nm} compared to \SI{532}{\nm}). Both our \SI{532}{\nm} system and the expected \SI{785}{\nm} system utilize the same Si avalanche photodiode technology. Analyzing our experimental data possessing these theoretical losses is therefore a valid proof-of-concept demonstration.

The experimental data are smoothed by taking the median of a moving window of \SI{29}{\s} width, the result illustrated in \cref{fig.varying_loss}. We use these smoothed data to select \SI{1}{\s} experimental data blocks to include in our analysis for each orbit by progressively scanning (from the center, in either direction) in \SI{1}{\s} steps for the next \SI{1}{\s} data block that possesses smoothed loss matching or exceeding the theoretical orbit loss prediction. By selecting experimental data at points where the smoothed loss is matched to theoretical link predictions, we ensure that the data we sample are not biased by normal fluctuations in measured loss.

\begin{figure}[tbp]
  \centering
  \includegraphics{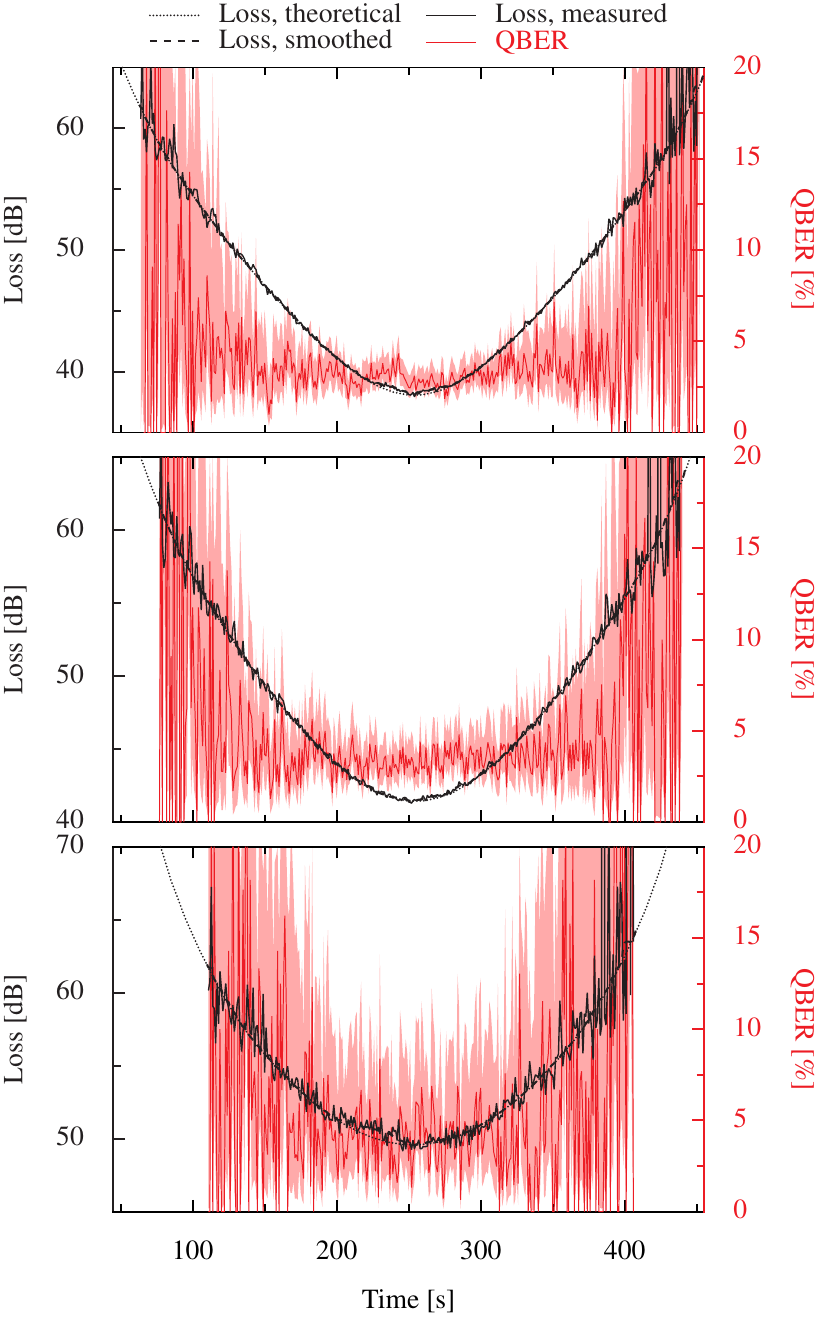}
  \caption{QBER and total loss of data sets reconstructed from measured data (shown in \cref{fig.finite_time}) for three representative satellite pass conditions: best pass (top), upper-quartile pass (middle), and median pass (bottom). The predicted loss is based on an uplink with a \SI{600}{\km} circular Sun-synchronous low Earth orbit satellite at a wavelength of \SI{785}{\nm}, with a receiver diameter of \SI{30}{\cm}, a \SI{2}{\micro rad} pointing error and a rural sea-level atmosphere. Smoothed loss follows the moving median determined at each \SI{1}{\s} experimental data block selected. Measured loss and QBER derive from the selected data, with shaded regions indicating the QBER \SI{95}{\percent} credible interval based on a uniform Bayesian prior. For the best pass, we obtain \SI{3374}{\bit} of secure key, including finite-size statistical effects.}
  \label{fig.loss_curve}
\end{figure}

\Cref{fig.loss_curve} shows the three relevant losses---the theoretically predicted loss, the smoothed loss value at the sampled point, and the experimentally measured loss from the sampled point---and the estimated QBER for each representative pass. The measured losses of the sampled experimental data closely match the trend of the theoretical prediction, whilst maintaining realistic fluctuation. At higher losses the per-second QBER estimate has significant fluctuations due to the reduced sample size.

Performing the post-processing steps on these data sets and incorporating finite-sized statistics, we are able to extract a \SI{3374}{\bit} secure key from the best pass, out of a total of \SI{544056}{\bit} raw key (\num{643521} detection events) with an average of \SI{3.1}{\%} QBER in the signal. This result shows that even with our modest \SI{76}{\MHz} source a positive key rate can feasibly be generated from one pass (albeit a good one) of a typical low Earth orbit satellite receiver. In comparison, the upper-quartile pass receives \SI{279317}{\bit} raw key (\num{348896} detections) with an average of \SI{3.5}{\%} QBER, but this is insufficient to produce nonzero secure key with finite-sized effects considered (the asymptotic secure key is \SI{17916}{\bit}). Similarly, the median pass with \SI{43375}{\bit} raw key (\num{82470} detections) and average \SI{4.4}{\%} QBER also cannot extract nonzero finite-sized secure key (asymptotic, \SI{877}{\bit}).

Improvements to the source could mitigate finite-sized statistical effects. By adjusting the photon and pulse count parameters, we can predict the performance of a \SI{400}{\MHz} source that produces $\sfrac{3}{4}$ signal (i.e.\ a \SI{300}{\MHz} signal rate, as per~\cite{Bourgoin2013}), $\sfrac{1}{8}$ decoy, and $\sfrac{1}{8}$ vacuum pulses. With other measured parameters left unchanged, \SI{21570}{\bit} secure key could be extracted from a single upper-quartile pass. This is directly comparable to the estimation of~\cite{Bourgoin2013}---under better conditions (e.g.\ $E_\nu$ assumed equal to $E_\mu$, intrinsic source QBER of 1\%, $\nu=0.1$, $f_\text{EC}=1.22$, and background estimate used only in calculation of $E_\mu$ while setting $Y_0=0$), a \SI{111.3}{\kilo\bit} secure key was predicted.

Alternatively, finite-size effects could be reduced by combining the measurements of multiple passes. For example, we are able to combine the measurements of three upper-quartile passes---each independently unable to produce positive finite key---and thereby extract \SI{165}{\bit} of secure key with finite-sized statistics. (Significantly more median passes, around 215 combined, would be required to yield positive finite-size secure key.) This might be a useful method to extract longer secure keys from the results of multiple marginal or individually unfruitful satellite passes.

\begin{table*}[t]
  \caption{Experimentally measured quantities for various loss conditions, including constant and varying losses simulating a satellite pass. Loss includes both channel loss and receiver efficiency. Except for rates extrapolated to a \SI{300}{\MHz} signal rate as indicated, all parameters are based on measurements with our \SI{76}{\MHz} pulsing laser source. Values here are incorporated into \cref{eq:SecureRate} and \cref{eq:FiniteSizeRate} to determine the appropriate size of the privacy amplification matrix (\cref{sec:PA}), and thus the final secure key length. Where necessary for the finite-size heuristic, uncertainties ($1\sigma$) are also given.}
  \centering
    \begin{tabular}{l|ccccccc|ccc}
    \hline\hline
	{\textbf{Loss [dB]}} & {\textbf{28.8}} & {\textbf{34.9}} & {\textbf{40.1}} & {\textbf{45.6}} & {\textbf{50.3}} & {\textbf{52.1}} & {\textbf{56.5}} & {\textbf{Best}}& {\textbf{Upper-quartile}} & {\textbf{Median}} \\
\hline
Duration [s] & 289 & 606 & 599 & 593 & 606 & 682 & 257 & 390 & 365 & 297 \\
Mean detection rate [\si{Hz}] & \num{41926} & \num{10464} & 3349 & 1167 & 427 & 301 & 186 & 1650 & 956 & 278 \\
Signal detections ($N_\mu$) [${\times}10^3$] & \num{11043} & 5739 & 1746 & 556 & 184 & 116 & 11.4 & 544 & 279 & 43.4 \\
Decoy detections ($N_\nu$) [${\times}10^3$] & 82.5 & 57.2 & 16.1 & 6.72 & 1.98 & 1.19 & 0.156 & 5.63 & 2.88 & 0.489 \\
Vacuum detections ($N_0$) [${\times}10^3$] & 43.8 & 34.2 & 14.4 & 7.75 & 5.30 & 4.19 & 0.612 & 4.64 & 3.32 & 1.50 \\
Signal photon number ($\mu$) & 0.506 & 0.490 & 0.507 & 0.579 & 0.534 & 0.503 & 0.581 & 0.505 & 0.507 & 0.512 \\
Decoy photon number ($\nu$) & 0.0392 & 0.0419 & 0.0515 & 0.0723 & 0.0517 & 0.0486 & 0.0592 & 0.0568 & 0.0571 & 0.0507 \\
Signal QBER ($E_\mu$) [\%] & 3.54 & 1.94 & 2.53 & 2.84 & 4.85 & 6.06 & 5.98 & 3.12 & 3.46 & 4.35 \\
\quad Uncertainty ($\sigma$) [${\times}10^{-2}$\,\%] & 0.566 & 0.581 & 1.20 & 2.26 & 5.14 & 7.24 & 22.4 & 2.15 & 3.52 & 10.0 \\
Decoy QBER ($E_\nu$) [\%] & 38.8 & 13.0 & 19.0 & 7.28 & 11.5 & 14.9 & 23.9 & 14.1 & 14.2 & 17.7 \\
\quad Uncertainty ($\sigma$) [${\times}10^{-1}$\,\%] & 2.17 & 1.51 & 3.44 & 3.29 & 7.62 & 11.2 & 39.1 & 4.55 & 7.04 & 19.0 \\
Signal vacuum QBER ($E_0^\mu$) [\%] & 50.8 & 52.0 & 50.4 & 50.6 & 50.3 & 50.4 & 50.5 & 50.6 & 50.7 & 50.7 \\
Decoy vacuum QBER ($E_0^\nu$) [\%] & 42.0 & 32.6 & 42.5 & 44.7 & 47.4 & 47.2 & 48.2 & 45.9 & 45.5 & 44.6 \\
Signal gain ($Q_\mu$) [${\times}10^{-6}$] & 568 & 136 & 41.8 & 13.5 & 4.34 & 2.43 & 0.634 & 20.1 & 11.0 & 2.10 \\
\quad Uncertainty ($\sigma$) [${\times}10^{-8}$] & 17.1 & 5.67 & 3.16 & 1.81 & 1.01 & 0.715 & 0.595 & 3.00 & 2.08 & 1.01 \\
Decoy gain ($Q_\nu$) [${\times}10^{-7}$] & 496 & 158 & 45.0 & 19.0 & 5.48 & 2.92 & 1.02 & 24.3 & 13.3 & 2.77 \\
\quad Uncertainty ($\sigma$) [${\times}10^{-8}$] & 17.3 & 6.61 & 3.55 & 2.32 & 1.23 & 0.848 & 0.817 & 3.54 & 2.47 & 1.25 \\
Single photon gain ($Q_1^\text{L}$) [${\times}10^{-6}$] & 370 & 111 & 24.5 & 7.70 & 2.65 & 1.31 & 0.400 & 12.1 & 6.35 & 1.10 \\
Single photon QBER ($E_1^\text{U}$) [\%] & 5.26 & 2.16 & 3.93 & 4.21 & 2.75 & 3.59 & 7.27 & 4.80 & 5.42 & 6.51 \\
Vacuum yield ($Y_0$) [${\times}10^{-7}$] & 20.6 & 7.39 & 3.14 & 1.71 & 1.15 & 0.806 & 0.312 & 1.56 & 1.19 & 0.665 \\
\quad Uncertainty ($\sigma$) [${\times}10^{-9}$] & 9.85 & 4.00 & 2.62 & 1.94 & 1.58 & 1.25 & 1.26 & 2.41 & 2.07 & 1.72 \\
Error correction efficiency ($f_\text{EC}$) & 1.41 & 1.50 & 1.40 & 1.35 & 1.17 & 1.12 & 1.13 & 1.26 & 1.223 & 1.15 \\
Raw rate [bits/s] & \num{38211} & \num{9470} & \num{2915} & 938 & 303 & 169 & 44.2 & \num{1395} & 765 & 146 \\
Sifted rate [bits/s] & \num{19298} & \num{3802} & \num{1447} & 469 & 150 & 84.1 & 21.7 & 694 & 379 & 71.6 \\
Secure rate (asymptotic) [bits/s] & \num{2684} & \num{1761} & 285 & 79 & 24.5 & 3.95 & 0.510 & 120 & 49.1 & 2.96 \\
Secure rate (finite-size) [bits/s] & \num{1935} & \num{1539} & 152 & 5.39 & -- & -- & -- & 8.65 & -- & -- \\
\quad At \SI{300}{\MHz}, projected [bits/s] & \num{12806} & \num{8683} & 1190 & 234 & -- & -- & -- & 372 & 59.1 & -- \\
Total finite-size key [kbits] & 559 & 932 & 91.2 & 3.20 & -- & -- & -- & 3.37 & -- & -- \\
\quad At \SI{300}{\MHz}, projected [kbits] & \num{3701} & \num{5262} & 713 & 138 & -- & -- & -- & 145 & 21.6 & -- \\
    \hline
    \end{tabular}

  \label{tab:Measured_quantities}
\end{table*}

The quantities measured in the experiment are summarized, in \cref{tab:Measured_quantities}, for each of the fixed loss cases and for each of the three varying-loss satellite-pass simulations. These are the values we use in \cref{eq:SecureRate} and \cref{eq:FiniteSizeRate} to determine the secure key length.

\section{Discussion and conclusion}\label{sec:Conclusion}

We have demonstrated the feasibility of satellite QKD using a quantum optical uplink by successfully performing QKD at losses up to \SI{56.5}{\dB} in the laboratory, with reduced computational requirements at the receiver, compatible with those that can be achieved on a satellite platform. We have improved over a previous high-loss demonstration~\cite{MeyerScott2011} by implementing complete QKD protocols, including twin-basis measurements, error correction and privacy amplification. We have also considered the effect of statistical fluctuations on the finite key length and have shown, by successfully performing full QKD and by extrapolation with varying losses that match those that would be experienced during representative passes of a satellite, that such a system is viable.

Several improvements to our system are possible as a next step, improving the key rate and moving our system towards being immediately deployable. One necessary modification to ensure secure QKD would be to employ a truly random source at Alice, rather than a fixed-length repeating pseudorandom sequence. Suitable high-speed electronics to implement this in tandem with a source with increased pulse rate could provide true security while significantly improving key rates above those reported here. While increased rates would necessitate more processing at the receiver, our analysis of computational requirements shows that detection rates could be increased by an order of magnitude or more over the demonstrated best-pass rate with processing at the receiver remaining feasible.

A particularly important challenge of satellite QKD yet to be addressed is the varying time of flight due to the changing distance between the satellite and ground station during a pass. For a \SI{600}{\km} orbit the distance between the satellite and ground station will vary by up to \SI{7}{\km/\s}~\cite{Bourgoin2013}, leading to a time of flight varying by up to \SI{23}{\micro\s} per second. Correcting for such variation as part of the timing analysis is straightforward, in principle, but is beyond the scope of the present work. Notably, though not of the same magnitude, varying time of flight due to relative motion has been demonstrated in the context of QKD for moving transmitters~\cite{Nauerth2012, Wang2013} and, very recently, a moving receiver~\cite{Bourgoin2015}.

Additionally, our theoretical prediction of loss for a satellite pass is based on a \SI{30}{\cm} diameter receiver at \SI{785}{\nm}, while our system uses a \SI{5}{\cm} receiver and operates at \SI{532}{\nm}. These differences do not affect the proof-of-concept demonstrated in this paper nor the basic design of our apparatus as the operating principles are the same in either case. However, the optimal parameters will need to be satisfied to ensure success of a satellite uplink---the increased telescope diameter is necessary to reduce the geometric losses, and the \SI{785}{\nm} wavelength is necessary to provide the best balance between diffraction, atmospheric absorption and turbulence, and detector efficiency~\cite{Bourgoin2013}. Together with a sufficiently accurate pointing mechanism, these engineering challenges for implementing a quantum receiver satellite payload are manifestly achievable in the near term.

\section{Acknowledgments}

We thank Chris Erven for valuable input and NSERC, Canadian Space Agency, CFI, CIFAR, Industry Canada, FedDev Ontario and Ontario Research Fund for funding. B.L.H. acknowledges support from NSERC Banting Postdoctoral Fellowships.

\end{document}